# Jointly Modeling Intra- and Inter-transaction Depend -encies with Hierarchical Attentive Transaction Embeddings for Next-item Recommendation

Shoujin Wang, Longbing Cao, *Senior Member, IEEE,* Liang Hu, Shlomo Berkovsky,
Xiaoshui Huang, Lin Xiao, Wenpeng Lu

**Abstract**—A transaction-based recommender system (TBRS) aims to predict the next item by modeling dependencies in transactional data. Generally, two kinds of dependencies considered are intra-transaction dependency and inter-transaction dependency. Most existing TBRSs recommend next item by only modeling the intra-transaction dependency within the current transaction while ignoring inter-transaction dependency with recent transactions that may also affect the next item. However, as not all recent transactions are relevant to the current and next items, the relevant ones should be identified and prioritized. In this paper, we propose a novel *hierarchical attentive transaction embedding* (HATE) model to tackle these issues. Specifically, a two-level attention mechanism integrates both item embedding and transaction embedding to build an attentive context representation that incorporates both intraand inter-transaction dependencies. With the learned context representation, HATE then recommends the next item. Experimental evaluations on two real-world transaction datasets show that HATE significantly outperforms the state-of-the-art methods in terms of recommendation accuracy.

✦

## 1 INTRODUCTION

GIVEN a transactional context, which consists of a set of recent transactions together with several existing items in the current transaction, a transaction-based recommender system (TBRS) aims to predict the next item that a user is likely to choose. It is usually formalized as a transaction-based next-item recommendation problem [1]. The set of recent transactions is treated as the inter-transaction context while the already-chosen items in the current transaction form the intra-transaction context. Generally speaking, the main challenge of next-item recommendations is to comprehensively capture the complex coupling relationships and interactions [2] embedded in the transactional data. In this work, we focus on dependency, which can be categorized into the *intratransaction dependency* between the intra-transaction context and the target items and the *inter-transaction dependency* between the inter-transaction context and the current transaction.

In the transactional data example shown in Fig. 1, a user has two recent transactions $t_1$ and $t_2$ and the current transaction $t_3$. We consider *milk* from $t_3$ as the target to recommend and all other prior transaction information as the corresponding context. Existing transaction-based next-item recommender systems (RSs) may suggest *salad* by only considering the intra-transaction items *apple* and *orange* in $t_3$, which may not be accurate as *salad* was just bought in $t_2$. Moreover, from the intra-transaction perspective, the choice of *milk* may depend much more on *bread* than on *apple* and *orange*. In such a case, a TBRS should be able to pay more attention to *bread* when modeling intra-transaction dependency. From the inter-transaction perspective, *milk* may also be influenced by *cake* and *egg* bought in $t_1$ but less related to $t_2$. This indicates that a good TBRS should not only take $t_1$ and $t_2$ into account but also concentrate much more on $t_1$. This example shows the importance of inter-transaction dependency and the significance of discriminating the contribution scales of different items and transactions according to their relevance to the next chosen item.

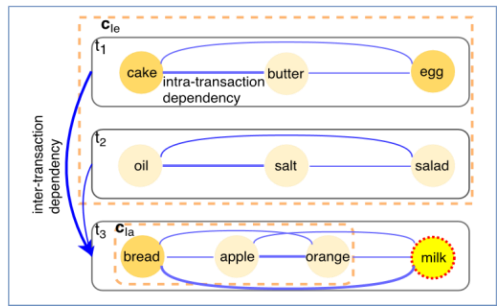

Fig. 1: Example of shopping transactions. Thicker lines and darker circles indicate stronger dependencies and the items more relevant to *milk*, while $c_{ia}(c_{te})$ represents the intra (inter)-transaction context.

Different approaches have been proposed to model the transaction dependencies for next-item recommendations. Pattern-based RSs predict the next item by using mined frequent patterns. Although easy to implement, the "support" constraint filters out many infrequent but interesting items and thus lead to information loss. Markov chain (MC) is an alternative way, but it only captures the first-order dependency between items [3]. To capture higher order dependency in sequential data, recurrent neural networks (RNN) have been successfully applied [4]. But the high computational cost caused by their complex structure prevents application to large data. Moreover, both MC and RNN assume a rigid order of items and thus the next choice is assumed to depend more on the recent items. Therefore, those truly relevant contextual items may not be paid enough attention to. To address such issues, researchers have incorporated the attention mechanism into RNN [5] or embedding model [6]. However, all these approaches only capture intra-transaction dependency while ignoring the rich intertransaction one, which may impact the next item, especially for periodic transactions. Although a variety of efforts have been made to incorporate the cross-transaction dependency with a hierarchical network structure, they either partly ignored the intertransaction transition relations by simply merging all historical transactions together to form one long-term set [7], or brought false dependency by modelling all historical transactions as a rigidly-ordered sequence [8], which may not always be the realworld case. Instead, not all recent transactions relate to the next choice, a priority should be given to those truly-related ones.

This paper addresses the above issues by proposing a novel *hierarchical attentive transaction embedding* (HATE) model. HATE first builds an attentive embedding for each transaction by emphasizing the relevant items in it and then builds attentive inter-transaction context embedding by highlighting those recent transactions more related to the current transaction and the next choice without rigid order assumption both within and between transactions. Simultaneously, an attentive intra-transaction context embedding is built on the items chosen in the current transaction. Finally, a hybrid context representation is achieved by combining both inter- and intra-transaction context embedding for the nextitem prediction.

Considering the large number of items in real-world data, it turns out to be practical to incorporate the attention mechanism into a shallow network in building a concise but powerful structure for attentive context representation learning. As a result, the proposed model is capable of capturing both intra- and intertransaction dependency attentively and the resultant context representation is more informative to predict the next item. Our validation on two real-world transaction datasets shows the necessity of combining the inter-transaction dependency with the attention mechanism. Accordingly, major contributions include:

- A hierarchical attentive transaction embedding model is proposed to learn the context representation for transaction-based item recommendations by attentively capturing both intra- and inter-transaction dependencies.
- A shallow and wide network is designed for efficiently learning the context representation over a large number of items and transactions.

In summary, our model relaxes the rigid order assumption both over items within a transaction and over transactions, which matches the real-word cases better. Empirical evaluation shows that (1) HATE outperforms the state-of-the-art TBRSs on realworld datasets by around 5%; (2) the incorporation of intertransaction context or attention mechanism achieves at least 10% accuracy improvement.

## 2 RELATED WORK

Rule- and pattern-based RSs are well-studied recommendation approaches [9]. To capture the transition between a sequence of songs, [10] discovered sequential patterns for next-song recommendations. Although being simple and effective, these methods often lose infrequent items [11]. More importantly, they only capture the co-occurrence relationships within transactions while ignoring the available inter-transaction dependency.

Markov chain (MC) models offer another way to model interitem transitions. Personalized Markov Embedding (PME) generates the embeddings of users and items in an Euclidean space for next-song recommendations [12]. Recently, to learn users' personalized sequential check-in information, a personalized ranking metric embedding method (PRME) was proposed for next POI recommendations [13]. Both PME and PRME are first-order MC models while the higher-order dependencies are ignored and the rigid order assumption over data may not always be realistic. More importantly, they are limited to the intra-transaction relations only, neglecting the inter-transaction dependency, which may lead to unreliable recommendations.

Recently, heuristics-based nearest neighbor (KNN) model was employed for session-based recommendations. Both item-based KNN [14] and session-based KNN [15] are proposed to model the intra- and inter-session dependencies respectively. Generally, it is good at capturing the natural co-occurrence based relations between items and the similarity relations between sessions. However, they lose the sequential dependencies over items and sessions and they treat all items or sessions equally important. Therefore, they cannot effectively emphasize those important items or transactions for next item recommendations.

RNN is a good choice to capture the higher-order dependency in TBRSs. Gated recurrent unit (GRU)-based RNN was proposed to capture long term dependency within [4], [16] or between transactions [17] , while hierarchical RNN [18] models were developed to capture the sequential dependency both within and between transactions. All these approaches model the intra- or intertransaction dependencies with a rigid order assumption, which may violate the real-world case since the transaction behaviours usually involve uncertainty and do not always follow a rigid order.

Recently, some researchers have introduced attention mechanisms into recommender systems to emphasize relevant and important information. Specifically, [5] incorporated attention mechanism into RNN to highlight those more important time-steps when modeling intra-transaction dependencies, while [7] proposed a hierarchical attention model to emphasize the relevant items from both the current short-term set and the historical long-term set. However, they still can not well capture the comprehensive intra- and inter-transaction dependencies in most real-world cases where both the order within and between transactions are relaxed rather than rigid. To substantially address such issue, we propose a hierarchical attentive transaction embedding model to learn a context representation by attentively capturing both intra- and inter-transaction dependencies for next-item recommendations.

## 3 PROBLEM STATEMENT

Given a transaction dataset, let $T = \{t_1, t_2 \dots t_{|T|}\}$ be the set of all transactions, such that each transaction $t = \{i_1, i_2 \dots i_{|t|}\}$ consists of a subset of items and is associated with a given user and a specified timestamp, where $|T|$ denotes the number of transactions in $T$. All the items occurring in all transactions constitute the whole item set $I = \{i_1, i_2 \dots i_{|I|}\}$. Note that the items in a transaction $t$ may not have a rigid order.

Given a target item $i_s \in t_j (j \cdot 6= 1)$, all other items in $t_j$ form the intra-transaction context $c_{Ia} = t_j \backslash i_s$. The recent transactions from the same user that happened before $t_j$ form the intertransaction context $c_{Ie} = \{t_1, t_2 \dots t_{j-1}\}$. $c_{Ia}$ and $c_{Ie}$ together constitute the transactional context $c = \{c_{Ia}, c_{Ie}\}$. Given the context $c$, HATE is trained as a probabilistic classifier that learns to predict a conditional probability distribution $P(i_s | c)$. Therefore, TBRS aims to rank all candidate items in terms of their conditional probability over the given context.

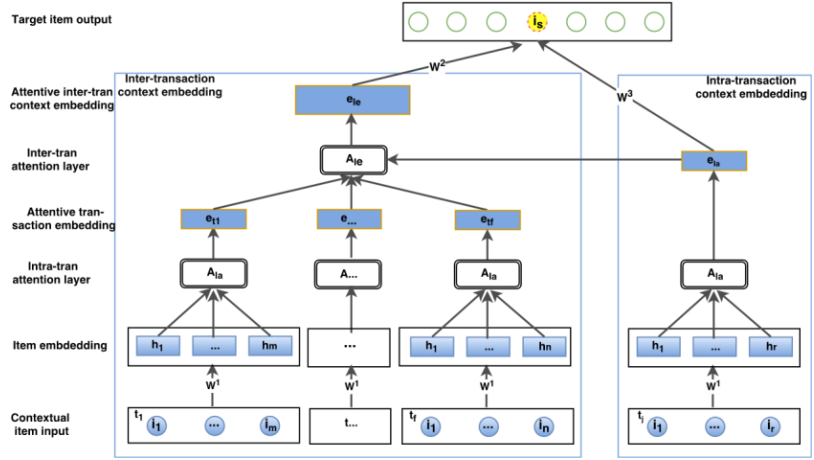

Fig. 2: The HATE architecture: It first learns item embedding, then integrates the embedding into intra-transaction context embedding or transaction embedding on which inter-transaction context embedding is learned. Both intra- and inter-transaction embedding is fed into the output layer for target item prediction. $A_{Ia}(A_{Ie})$ represents the intra- and inter-transaction attention model.

## 4 MODELING AND LEARNING

### 4.1 The HATE Model

As shown in Figure 2, the proposed HATE model consists of two main parts: the transactional context embedding part at the bottom and the prediction part (output layer) at the top. The embedding part contains two modules: inter-transaction context embedding and intra-transaction context embedding.

#### 4.1.1 Inter-transaction Context Embedding

Item embedding. For a given contextual item $i_l$ from a transaction $t$, we create an embedding mechanism to map its ID number to an informative and low-dimensional vector representation in the item embedding layer, where a K-dimensional real-valued vector $\mathbf{h}_l \in R^K$ is used to represent item $i_l$. The input weight matrix $\mathbf{W}^1 \in R^{K \times |l|}$ is used to fully connect the input-layer and item embedding-layer. Note that actually the $l^{th}$ column of $\mathbf{W}^1$ encodes item $i_l$ to a real-valued embedding $\mathbf{h}_l$ as below. Several different mapping approaches including logistic function have been tried to map item ID to its embedding and the following way is found to achieve the best performance in our case.

$$\mathbf{h}_l = \mathbf{W}^1_{:,l} \tag{1}$$

Attentive transaction embedding. When the embeddings of all the items in transaction $t$ are ready, we can obtain the embedding $\mathbf{e}_t \in R^K$ of contextual transaction $t$ by integrating the embeddings of all items in $t$ using the attention mechanism. Specifically, the attentive transaction embedding is built as a weighted sum of $\mathbf{h}_l$:

$$\mathbf{e}_t = \sum_{i_l \in t} \alpha_{sl} \mathbf{h}_l, \ s.t. \sum_{i_l \in t} \alpha_{sl} = 1 \tag{2}$$

where $\alpha_{sl}$ is the integration weight of contextual item $i_l$ w.r.t. the target item $i_s$, indicating the contribution scale of $i_l$ to the choice of $i_s$. In our model, to better capture the different contribution scales of contextual items, we develop an attention layer to learn the integration weights automatically and effectively. Compared with assigning the weights manually under certain assumptions, e.g., order assumption, or directly learning the weights without the attention mechanism, our method not only works more flexibly without assumptions but also emphasizes those important items and reduces the interference from irrelevant ones. Next, we demonstrate how the intra-transaction attention model achieves this goal.

Intra-transaction attention. Similar to most attention models, we use a softmax layer to learn the weights of different contextual items w.r.t the target item. In this way, items that are more relevant to the target item are given larger weights, and vice versa. The input of softmax is the transformation of each item's embedding:

$$\alpha_{sl} = \frac{exp(\sigma(\mathbf{h}_l))}{\sum_{i_v \in t} exp(\sigma(\mathbf{h}_v))} \tag{3}$$

$$\sigma(\mathbf{h}_l) = \mathbf{w}^{aT}\mathbf{h}_l \tag{4}$$

where $\mathbf{w}^a$ is an item-level context vector shared by all contextual items, which can be seen as a high level representation of a fixed query "which item is relevant to the target item?" over all the contextual items. The vector is randomly initialized and jointly learned during the training stage. As $\mathbf{w}^a$ serves as a weight vector connecting the item embedding layer to the intratransaction attention model, we denote it as an intra-transaction attention weight, to be consistent with input and output weights. Essentially, the importance of each item $i_l$ is achieved by first calculating the similarity between its embedding $\mathbf{h}_l$ and the item level context vector $\mathbf{w}^a$ and then normalizing it into an importance weight $\alpha_{sl}$ through a softmax function.

Attentive inter-transaction context embedding. Intertransaction context embedding is built on top of the embeddings of transactions included in the inter-transaction context. Specifically, the inter-transaction context embedding is computed as a weighted sum of transaction embeddings:

$$\mathbf{e}_{Ie} = \sum_{t_x \in c_{Ie}} \beta_{sx} \mathbf{e}_{t_x}, \ s.t. \sum_{t_x \in c_{Ie}} \beta_{sx} = 1 \tag{5}$$

where $\beta_{sx}$ is the integration weight of transaction $t_x$ from the inter-transaction context $c_{Ie}$ for the target item $i_s$. It indicates the relevance degree of $t_x$ to the current transaction, i.e., intratransaction context $c_{Ia}$, by modeling the interaction between $t_x$ and $c_{Ia}$ in the inter-transaction attention model $A_{Ie}$. More relevant to the current transaction, $t_x$ will be more influential on the choice of $i_s$, therefore $\beta_{sx}$ essentially implies the contribution scale of transaction $t_x$ to the choice of the target item $i_s$.

Inter-transaction attention. Differently from the intratransaction attention model, except for the transactions from intertransaction context, we take the intra-transaction context as an additional input to model the interaction between transactions as indicated in Figure 2. We first use a matrix to model the interactions between each inter-transaction and the intra-transaction context, and then import the product of inter-transaction embedding, interaction matrix and intra-transaction context embedding into the attention model.

$$\beta_{sx} = \frac{exp(\varrho(\mathbf{e}_{t_x}))}{\sum_{t_f \in \mathbf{c}_{Ie}} exp(\varrho(\mathbf{e}_{t_f}))} \tag{6}$$

$$\varrho(\mathbf{e}_{t_x}) = \mathbf{e}_{t_x}^T \mathbf{W}^\beta \mathbf{e}_{Ia} \tag{7}$$

where $\mathbf{W}^\beta$ is a transaction-level interaction matrix shared by all the contextual transactions. It can be regarded as a high level representation of a query "which transaction in the intertransaction context is relevant to the current one?". This matrix is randomly initialized and jointly learned during the training process. We refer it to as the inter-transaction attention weight. $\mathbf{e}_{Ia}$ is the embedding of intra-transaction context and its calculation will be given shortly.

#### 4.1.2 Intra-transaction Context Embedding

Given an intra-transaction context $c_{Ia}$ consisting of multiple chosen items in the current transaction, we first get the embedding of each item with the aforementioned item embedding. Then we integrate these embeddings attentively to build the intratransaction context embedding.

$$\mathbf{e}_{Ia} = \sum_{i_z \in \mathbf{c}_{Ia}} \alpha_{sz} \mathbf{h}_z, \qquad s.t. \sum_{i_z \in \mathbf{c}_{Ia}} \alpha_{sz} = 1 \tag{8}$$

where $h_z$ is the embedding of an intra-transaction context item $i_z$ and is calculated using Equation (1) while $\alpha_{sz}$ is the integration weight calculated using Equations (3) and (4).

#### 4.1.3 Target Item Prediction

Once the embeddings of both intra- and inter-transaction contexts are ready, we feed them into the output layer for the target item prediction, as shown in the upper part of Figure 2. Here the output weight matrix $\mathbf{W}^2 \in \mathbb{R}^{|I| \times K}$ and $\mathbf{W}^3 \in \mathbb{R}^{|I| \times K}$ are used to fully connect the intra- and inter-transaction context embeddings to the output layer. Specifically, given the context embeddings and the weights, a score indicating the possibility of the choice of a target item $i_s$ under the context c is computed using:

$$S_{i_s}(\mathbf{c}) = \mathbf{W}_{s,:}^2 \mathbf{e}_{Ie} + \mathbf{W}_{s,:}^3 \mathbf{e}_{Ia} \tag{9}$$

where $\mathbf{W}_{s,:}^2$ denotes the $s^{th}$ row of $\mathbf{W}^2$ and $S_{i_s}(\mathbf{c})$ quantifies the relevance of the target item $i_s$ w.r.t. the given context c. Therefore, the conditional probability distribution $P_\Theta(i_s|c)$ is defined with the commonly used softmax function:

$$P_\Theta(i_s|\mathbf{c}) = \frac{exp(S_{i_s}(\mathbf{c}))}{Z(\mathbf{c})} \tag{10}$$

where $Z(\mathbf{c}) = \sum_{i \in I} exp(S_i(\mathbf{c}))$ is the normalization constant and $\Theta = \{\mathbf{W}^1, \mathbf{w}^\alpha, \mathbf{W}^\beta, \mathbf{W}^2, \mathbf{W}^3\}$ includes the model parameters. Therefore, a probabilistic classifier modeled by the proposed HATE model is obtained to predict the target item and accordingly recommend the next item.

### 4.2 Parameter Learning and Item Prediction

We now discuss how to learn the model parameters and predict the next item using the trained model in this section.

A probabilistic classifier is built over the transaction data $d = $hc,$i_c$i, where c is the input context and $i_c$ is the observed output conditional on c. Given a training dataset $D = \{$hc,$i_c$i$\}$, the joint probability distribution is obtained by:

$$P_\Theta(D) \propto \prod_{d \in D} P_\Theta(i_c|\mathbf{c}) \tag{11}$$

Therefore, the model parameters $\Theta$ can be learned by maximizing the conditional log-likelihood (cf. Equation. (10)):

$$L_\Theta = \sum_{d \in D} log P_\Theta(i_c|\mathbf{c}) \tag{12}$$

Note that the evaluation of $L_\Theta$ and its corresponding gradient computation involve the normalization term $Z(\mathbf{c})$, the computation of which is time consuming as it sums $exp(S_{i_c}(\mathbf{c}))$ over all the items for each training instance. The commonly used noisecontrastive estimation (NCE) technique [19] is adopted here to enhance the training efficiency. NCE uses a binary classifier to distinguish samples from the data distribution from those with a known noise distribution to avoid the high computation cost when computing the normalization constant of the softmax.

Once the model parameters $\Theta$ have been learned, HATE is ready to compute predictions and thus generate next-item recommendations. Specifically, given an arbitrary transactional context which contains both intra- and inter-transaction contexts indicating prior transaction data of a user, the probabilities of choosing next candidate items are calculated according to Equation (10), and a ranking reflecting the priority of the candidate items is achieved.

## 5 EXPERIMENTS AND EVALUATION

### 5.1 Experimental Setup

#### 5.1.1 Dataset Preparation

We evaluate our proposed method on two real-world grocery store transaction datasets: a public dataset Dunnhumby[1] and a proprietary Australian national supermarket (ANS) dataset [20]. Dunnhumby includes transaction records of around 2,500 households shopping frequently at multiple stores of the same retailer over two years. ANS contains transaction records of about 1,000 customers, collected by an Australian national supermarket chain within a period of one year.

First, a sequence of transactions is extracted for every user and then a sliding window is used to cut each user's transactions sequence into multiple triple-transaction units. For each unit, we consider the first two transactions as the inter-transaction context $c_{Ie}$ and the last one as the current transaction. Our selection is data-drive and is explained by the most frequently observed transaction pattern of three transactions per week in the shopping cycle. Each

                                                                 

time one item from the current transaction is picked up as the target item $i_s$ and all others are considered as the intratransaction context $c_{ta}$. We do this because the order information over items within transactions is not provided and thus we relax the rigid order assumption. As a result, the training and test instances are built in the format of $d = $ hc,$i$,i(c = {$c_{to}, c_{ta}$}) as illustrated in the previous section. Finally, we randomly select 20% of transactions that occurred in the last 30 days as the test set and leave the remainder for training. The characteristics of the datasets are shown in Table 1.

TABLE 1: Statistics of experimental datasets

| Statistics | Dunnhumby | ANS |
|---|---|---|
| #Transactions | 65,001 | 99,987 |
| #Items | 10,292 | 11,996 |
| Avg. Transaction Length | 12.15 | 10.81 |
| #Training Sequence of Trans. | 149,606 | 258,561 |
| #Training Instances | 402,739 | 703,062 |
| #Test Sequence of Trans. | 7,874 | 13,608 |
| #Test Instances | 21,205 | 36,933 |

### 5.1.2 Comparison Methods and Metrics

We use the following methods as the evaluation baselines.

- **PBRS**: A typical pattern-based recommender which uses mined frequent patterns to generate recommendations [21].
- **FPMC**: A model that factorizes the personalized transition matrix between items with pairwise interactions for nextbasket recommendation [22].
- **PRME**: A personalized ranking metric embedding model (PRME) for next POI recommendations with a Markov chain framework [13].
- **GRU4Rec**: A typical session-based RS built on RNN. It models the session sequence using a GRU-based RNN framework [4].
- **SWIWO**: A shallow wide-in-wide-out network embedding model for session-based RSs [23].
- **NCSF**: An RNN-based neural architecture to model both intra- and inter-context for next item prediction [17].
- **SHAN**: A two-layer hierarchical attention network to learn both users' long- and short-term preferences for next item prediction [7].
- **ATE**: A model similar to HATE that only utilizes the intratransaction context. This assesses the contribution of the intertransaction context.
- **HTE**: A model similar to HATE that replaces the intertransaction attention module with a fully-connected layer. This assesses the effect of the inter-transaction attention module.

Two common accuracy metrics are used in the evaluation.

- **REC@K**: measures the recall of the top-K ranked items in the recommendation list. We choose $K \in \{10, 50\}$ as users are usually interested only in top items. Specifically, for N top-K recommendations, the corresponding **REC@K** is calculated:

$$REC@K = \frac{1}{N}\sum_{j=1}^{N}|R_j \cap i_{s_j}| \qquad (13)$$

where $R_j$ and $i_{s_j}$ are the $j^{th}$ recommendation list and the corresponding true next item respectively.

- **MRR**: measures the mean reciprocal rank of the predictive position of the true target item.

## 5.2 Performance Evaluation

### 5.2.1 Accuracy Evaluation

Table 2 and Table 3 show the obtained REC@10, REC@50 and MRR on two real-world transaction datasets. We empirically set the minimum support to 0.02 on both datasets in PBRS. The information loss caused by filtering out infrequent items leads to poor performance. To achieve the best performance, we set the factor number to 10 for FPMC which performs not good on both datasets, mainly caused by the data sparsity. Due to the large numbers of transactions and items but limited interactions between them, quite large but very sparse item transition matrices are constructed to train this MF model. Following [13], the embedding dimension is set to 60 for PRME. As a first-order MC model, PRME is easy to lose information by learning the transition probability over the successive item instead of the whole context. In addition, the rigid order assumption set by these models may not always match the real world purchasing events. GRU4Rec achieves much better performance compared to the above three methods by benefiting from its deep structure. Building a flexible embedding on the whole context, SWIWO is able to capture the complex intra-transaction dependency for better recommendations. A common drawback of all these models is that they are all limited to the intra-transaction dependency. In contrast, NCSF performs better by incorporating inter-transaction dependency for next-item prediction. However, it assumes a rigid order assumption over historical transactions by employing RNN, which may not be the case. SHAN attentively incorporate intertransaction dependencies and performs even better. But it breaks down the structures of long-term transactions and put all of their items into one pool, which may loss the intra- and inter-transaction dependencies embedded in long-term transactions and thus reduce the recommendation performance.

TABLE 2: Accuracy comparisons on Dunnhumby

| Model | REC@10 | REC@50 | MRR |
|---|---|---|---|
| *PBRS* | 0.0817 | 0.0901 | 0.0421 |
| *FPMC* | 0.0333 | 0.0711 | 0.0317 |
| *PRME* | 0.0757 | 0.0912 | 0.0613 |
| *GRU4Rec* | 0.2018 | 0.3002 | 0.1216 |
| *SWOWI* | 0.2469 | 0.3379 | 0.1139 |
| *NCSF* | 0.2769 | 0.3828 | 0.1284 |
| *SHAN* | 0.2908 | 0.4308 | 0.1346 |
| *HATE* | 0.3012 | 0.4513 | 0.1421 |
| *ATE* | 0.2752 | 0.3754 | 0.1250 |
| *HTE* | 0.2752 | 0.4000 | 0.1218 |

TABLE 3: Accuracy comparisons on ANS

| Model | REC@10 | REC@50 | MRR |
|---|---|---|---|
| *PBRS* | 0.0572 | 0.0765 | 0.0410 |
| *FPMC* | 0.0310 | 0.0555 | 0.0292 |
| *PRME* | 0.0611 | 0.0800 | 0.0522 |
| *GRU4Rec* | 0.1405 | 0.2951 | 0.0755 |
| *SWOWI* | 0.1400 | 0.3015 | 0.0805 |
| *NCSF* | 0.1501 | 0.3250 | 0.0895 |
| *SHAN* | 0.1616 | 0.3396 | 0.0932 |
| *HATE* | 0.1756 | 0.3515 | 0.0993 |
| *ATE* | 0.1542 | 0.2254 | 0.0805 |
| *HTE* | 0.1756 | 0.2755 | 0.0874 |

For our HATE model, the embedding dimension and the batch size are empirically set to 50 and 30 respectively on both datasets. Adagrad with an initial learning rate of 0.5 is applied to train the model. By attentively learning the hierarchical dependencies embedded in the inter-transaction context and then attentively combining it together the intra-transaction dependency for next-item prediction, HATE outperforms the best baseline SHAN by 4.64% and 6.24% in average on Dunnhumby and ANS respectively, which validates the advantage of our model. In particular, the enhanced 10% performance of HATE compared to ATE and HTE demonstrates the significance of incorporating inter-transaction context and attention mechanism respectively. Particularly, the hierarchical attention mechanism helps to emphasize those truly relevant items and transactions when modeling dependency. Note that a minority of true target items are ranked very high in our recommendation lists while some others are ranked very low, leading that even the MRR is larger than 0.1 (cf. Table 2) but the REC@10 is not so high as expected.

### 5.2.2 The Effect of Number of Incorporated Inter Transactions

Generally speaking, a long inter-transaction context which contains more recent transactions is more likely to include transactions irrelevant to the current transaction and the next-item choice. As a result, it is harder to identify and emphasize those truly relevant transactions in a long context. To show the advantage of attention mechanism in handling long contexts, we test the effect of the number of incorporated inter-transactions on a subset of Dunnhumby by selecting users with at least 6 transactions. Each time a different number of recent transactions is considered as the inter-transaction context. Figure 3 shows that HATE gains larger margins compared with others when incorporating more transactions, which demonstrates its ability to emphasize the relevant transactions in longer inter-transaction contexts. We compare here HATE and other three approaches because only these four approaches can incorporate inter-transaction context.

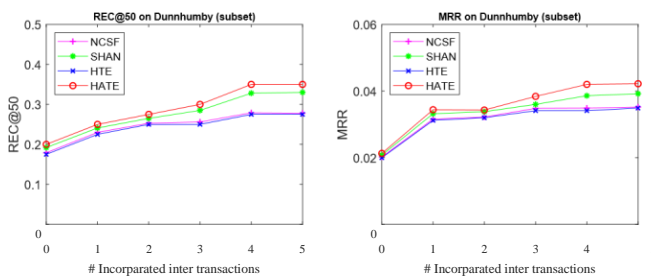

Fig. 3: HATE gains larger margins when incorporating more intertransactions.

## 6 Conclusions

This work proposes a hierarchical attentive transaction embedding model HATE - a shallow and wide neural network for transaction embedding. By incorporating both current transaction and recent transactions, HATE is able to capture both intra- and intertransaction dependencies and build a more informative context representation. In addition, the incorporation of hierarchical attention models allows us to emphasize items and transactions particularly relevant to the next-item choice when building the attentive representation, leading to better recommendation. Empirical validation on two real-world transaction datasets shows the superiority of HATE over several state-of-the-art approaches. We will explore the applications of HATE to other problems, e.g., document analysis and multimedia recommendations, and will learn more complex couplings and interactions in transactions.

## 7 Acknowledgement

This work is partially sponsored by the Australian Research Council Discovery grant DP190101079 and the ARC Future Fellowship grant FT190100734.

**Shoujin Wang** is now a research fellow of data science and artificial intelligence with Macquarie University, Sydney, Australia. His primary research interests include data mining, machine learning and recommender systems. This work was done during his PhD candidate with the University of Technology Sydney, Sydney, Australia. Contact him at: shoujin.wang@uts.edu.au.

**Longbing Cao** is a Professor of information technology with the University of Technology Sydney, Australia. His primary research interest includes data science, machine learning, behavior informatics, and their enterprise applications. Contact him at: longbing.cao@uts.edu.au.

**Liang Hu** is a research assistant with University of Technology Sydney, Sydney, Australia. His research interests include recommender systems, data mining, machine learning. Contact him at: rainmilk@gmail.com.

**Shlomo Berkovsky** is an associate Professor and the leader of the Precision Health research stream at the Australian Institute of Health Innovation, Macquarie University. Shlomo also studies how sensors and physiological responses can predict medical conditions, and how clinicians and patients interact with health technologies. His areas of expertise also include user modelling, online personalisation, and persuasive technologies. Contact him at: shlomo.berkovsky@mq.edu.au.

**Xiaoshui Huang** is a research associate with the University of Sydney, Sydney, Australia. His research interests include computer vision, digital health, machine learning. Contact him at: xiaoshui.huang@sydney.edu.au.

**Lin Xiao** is a professor with Hunan Normal University, Changsha, China. His main research interests include neural networks, robotics, and intelligent information processing. Contact him at: xiaolin860728@163.com.

**Wenpeng Lu** is an associate professor of computer science and technology with the Department of Computer Science and Technology, Qilu University of Technology (Shandong Academy of Sciences), Jinan, China. Contact him at: wenpeng.lu@qlu.edu.cn.